\documentclass{article}
\usepackage{spconf,amsmath,graphicx}
\usepackage{hyperref}
\usepackage{booktabs}
\usepackage{multirow}
\usepackage{color}
\usepackage{amssymb}
\usepackage{multicol}

\title{InstructSing: High-Fidelity Singing Voice Generation via Instructing Yourself}
\name{$^*$Chang Zeng$^{1}$, $^*$Chunhui Wang$^2$, Xiaoxiao Miao$^3$, Jian Zhao$^2$, Zhonglin Jiang$^2$, Yong Chen$^{2}$}
\address{$^1$National Insitute of Informatics, Japan \; $^2$Geely, China \\ $^3$Singapore Institute of Technology, Singapore}
\begin{document}
\ninept
\maketitle
\renewcommand{\thefootnote}{\fnsymbol{footnote}}
\footnotetext[1]{These authors contributed equally to this work.}
\renewcommand{\thefootnote}{\arabic{footnote}}
\begin{abstract}
It is challenging to accelerate the training process while ensuring both high-quality generated voices and acceptable inference speed. In this paper, we propose a novel neural vocoder called InstructSing, which can converge much faster compared with other neural vocoders while maintaining good performance by integrating differentiable digital signal processing and adversarial training. It includes one generator and two discriminators. Specifically, the generator incorporates a harmonic-plus-noise (HN) module to produce 8kHz audio as an instructive signal. Subsequently, the HN module is connected with an extended WaveNet by an UNet-based module, which transforms the output of the HN module to a latent variable sequence containing essential periodic and aperiodic information. In addition to the latent sequence, the extended WaveNet also takes the mel-spectrogram as input to generate 48kHz high-fidelity singing voices. In terms of discriminators, we combine a multi-period discriminator, as originally proposed in HiFiGAN, with a multi-resolution multi-band STFT discriminator. Notably, InstructSing achieves comparable voice quality to other neural vocoders but with only one-tenth of the training steps on a 4 NVIDIA V100 GPU machine\footnote{{Demo page: \href{https://wavelandspeech.github.io/instructsing/}{\texttt{https://wavelandspeech.github.io/inst\\ructsing/}}}}. We plan to open-source our code and pretrained model once the paper get accepted.
\end{abstract}

\section{Introduction}
\label{sec:intro}

Thanks to the development of deep learning, neural network-based audio processing has achieved great success in tasks of generating realistic and natural sound and voice \cite{fastspeech,xiaoice2,audioldm}. Among these tasks, singing voice synthesis (SVS) has attracted extensive attention from both the academy and industry \cite{refinegan,hifisinger,diffsinger}. The pipeline of a contemporary SVS system can be decomposed into two stages: 1) the first stage involves the acoustic model \cite{xiaoice2,xiaoice1,crosssinger,hono2023singing,Kakoulidis_2022}, primarily responsible for generating the intermediate representation, such as mel-spectrogram, based on the lyrics and MIDI information extracted from the musical score; 2) the second stage employs a vocoder \cite{yoneyama2023sourcefilter,kim2022adversarial,huang2021multisinger,hifiwavegan} to convert the intermediate representation, initially generated by the acoustic model, into an audible waveform. To achieve satisfactory results, both the acoustic model and the vocoder demand access to extensive high-quality singing data.

Different from the text-to-speech (TTS) task, the SVS task aims at generating high-fidelity singing voices with a higher sampling rate (e.g. 48kHz) for better auditory perception. To achieve this, many neural vocoders \cite{refinegan,hifiwavegan,usfgan,hn-usfgan} for the SVS task were improved based on their counterparts \cite{nsf,nsf2,pwg,kong2020hifigan} for the TTS task. For instance, 
the HiFi-WaveGAN model \cite{hifiwavegan} improves the architecture of the Parallel WaveGAN \cite{pwg} by increasing the size of the receptive field of the generator. Moreover, it creates a pulse sequence according to digital signal processing (DSP) to regulate the behavior of the generator to avoid distortion when generating the waveform. While these works have obtained good performance on the SVS task, they still suffer from the slow training speed for convergence. Additionally, there are some lightweight vocoders such as differentiable DSP (DDSP)-based models \cite{ddsp,lpcnet} that can converge with a fast training speed. However, their poor performance in singing voice generation cannot meet the needs of the SVS task.

Inspired by the advantages of the DDSP-based method, which offers fast training convergence, and the generative adversarial network (GAN)-based method known for its good performance, this work aims to achieve a balanced trade-off to expedite the training process while ensuring both high-quality generated voices and acceptable inference speed, thereby increasing productivity and efficiency. To achieve this goal, we propose a novel neural vocoder named InstructSing for SVS tasks by combining a harmonic-plus-noise (HN) module \cite{harmonic_noise1,harmonic_noise2} with an extended WaveNet (ExWaveNet) \cite{hifiwavegan} through an UNet-based \cite{ronneberger2015unet} module and leverage adversarial training to make full use of the merits of them. Specifically, the proposed model utilizes the HN module to generate the harmonic content and noise as an instructive signal sequence to guide the training of the rest modules. Therefore, we call this module InstructNet. Note here we apply an additional reverberation module \cite{ddsp} on the harmonic content and noise to generate 8kHz audio only at the training stage to meet the demand of backward propagation based on spectral loss \cite{nsf,nsf2,ddsp,li2023design}. Subsequently, this instructive signal sequence is transformed by an UNet module to a latent variable sequence, which contains sufficient periodic and aperiodic knowledge beneficial to the following 48kHz waveform generation process. The latent sequence plays a sine excitation-like \cite{huang2022singgan,nsf,nsf2,ddsp} role in generating audio, but we argue that knowledge contained in this sequence is superior compared with the pure DSP-based sine excitation because it has been refined via the UNet module. We call this UNet-based module BridgeNet since it is like a bridge that connects the InstructNet with the ExWaveNet, which is responsible for generating the high-fidelity 48kHz waveform from the mel-spectrogram and corresponding latent variable sequence. 

Additionally, we utilize a multi-period discriminator (MPD) \cite{kong2020hifigan} and a multi-resolution multi-band STFT discriminator (MR-MBSD) to further enhance audio quality from the time domain and frequency domain, respectively. The latter is improved from the multi-resolution STFT discriminator (MRSD) \cite{univnet} by incorporating the multi-band (MB) analysis suggested in \cite{multibandmelgan, chen2023gesper}, acknowledging that distinct subbands exhibit varying patterns. It builds upon the concept of multi-band analysis \cite{durian}, utilizing multiple equal divisions of STFT features to extend the MRSD to encompass multiple sub-bands.

\begin{figure*}[t]
  \centering
  \includegraphics[width=0.9\linewidth]{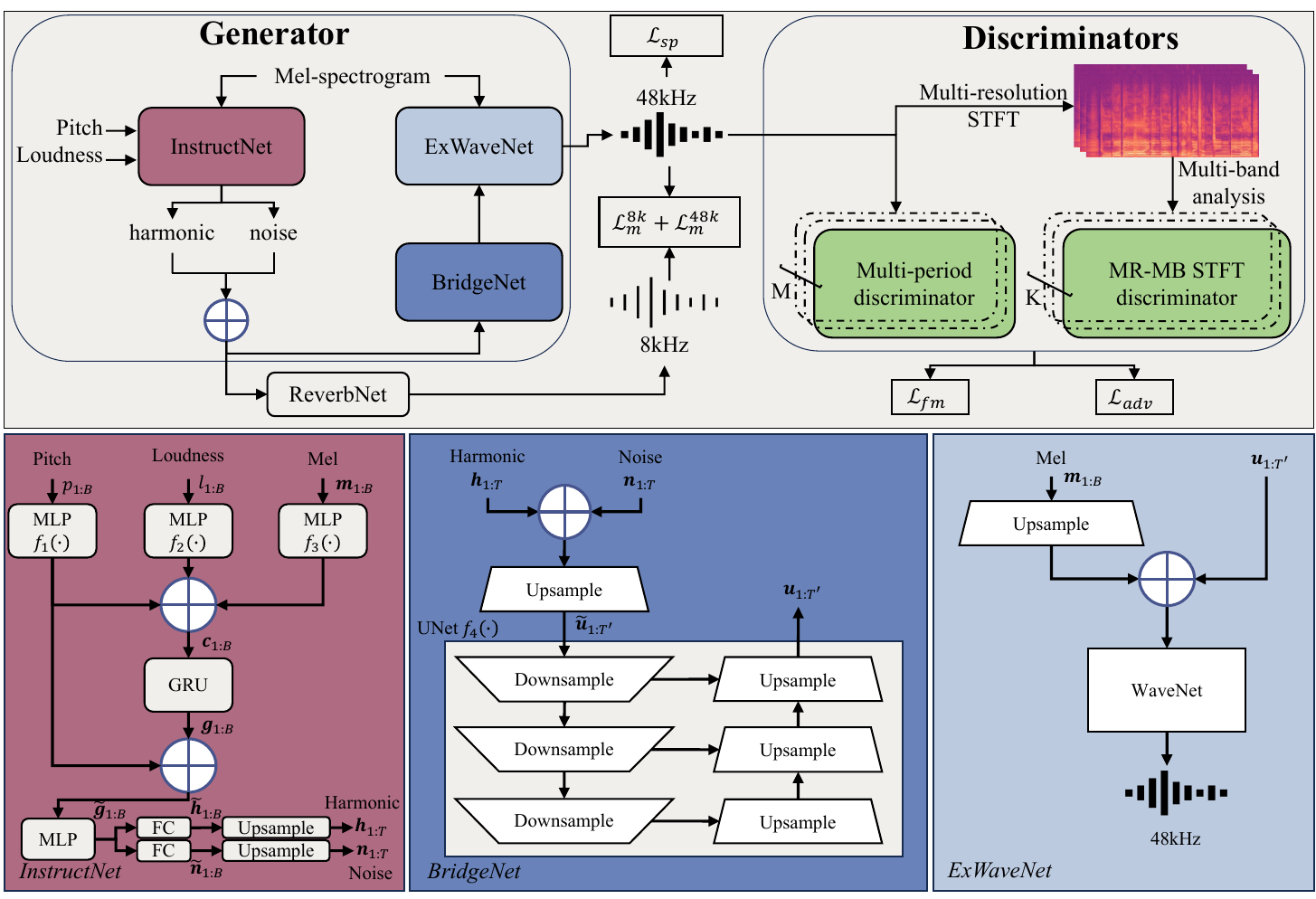}
  \caption{Architecture of the proposed InstructSing. It is a GAN-based model comprised of a generator and two distinct discriminators, along with a detailed structure of the generator, which includes an InstructNet, a BridgeNet, and an Extended WaveNet.}
  \label{fig:arch-instructsing}
\end{figure*}

In general, InstructSing offers several advantages in achieving a balance between training time and high-quality voices. The DDSP-based InstructNet, serving as one of the generator components, can generate harmonic and noise sequences. These sequences not only produce low-resolution 8kHz audio as instructive signals to accelerate model convergence but also provide enriched periodic and aperiodic knowledge as guidance. After transformation, they are fed to ExWaveNet, effectively eliminating glitches in the reconstruction of high-fidelity 48kHz audio and providing a clearer learning objective.
Furthermore, the MR-MBSD is employed to enhance audio quality through adversarial training. Experimental results demonstrate that InstructSing can converge within 20,000 training steps, which is only one-tenth of other strong baseline systems \cite{refinegan,hn-usfgan} when training on a 4 NVIDIA V100 GPU machine. It outperforms them in both training speed and voice quality, with acceptable inference speed.

The rest of this paper is organized as follows. 
Our proposed InstructSing is concretely illustrated in Section \ref{sec:instructsing}, including the generator, discriminators, and several objective functions. The experimental setup, as well as the results are shown in Section \ref{sec:exp}. Finally, we summarize our paper in Section \ref{sec:conclusion}.

\section{InstructSing}
\label{sec:instructsing}

Figure \ref{fig:arch-instructsing} (top) provides an overview of the proposed InstructSing, comprising a generator and two discriminators. The generator uses mel-spectrogram, pitch sequence, and loudness sequence inputs to create a 48kHz high-fidelity waveform. In terms of discriminators, the MPD \cite{kong2020hifigan} identifies crucial periodic patterns in the generated waveform from the time domain for better auditory perception. Meanwhile, the MR-MBSD distinguishes real and generated waveforms from the frequency domain rather than the time domain in \cite{kong2020hifigan,kumar2019melgan} by analyzing long-term dependencies with multiple STFT parameter sets \cite{univnet} and multi-band analysis \cite{multibandmelgan}. This section will delve into the intricacies of InstructSing.

\subsection{Generator}
\label{sec:generator}

InstructSing, being a GAN-based model akin to others such as \cite{hifiwavegan, kong2020hifigan}, possesses the capacity to produce high-quality waveforms through adversarial training. However, our paper targets accelerating neural vocoder convergence while upholding comparable performance. To achieve this, leveraging the quick convergence trait observed in DDSP-based models \cite{nsf,nsf2,ddsp} when generating audio with low sampling rate, we incorporate the DDSP-based InstructNet into the generator, as top left of Figure \ref{fig:arch-instructsing} shows, to produce the harmonic content and noise as an instructive signal for the following adversarial training. The instructive signal undergoes refinement by BridgeNet, generating a latent variable sequence rich in precise periodic and aperiodic information. This sequence significantly expedites the convergence of ExWaveNet \cite{hifiwavegan}, responsible for generating 48kHz high-fidelity singing voices.

\subsubsection{InstructNet}
\label{sec:instructnet}

Following the design of the harmonic-plus-noise model \cite{ddsp,harmonic_noise1,harmonic_noise2}, the structure of InstructNet is shown on the bottom left of Figure \ref{fig:arch-instructsing}.

Suppose the input mel-spectrogram is $\boldsymbol{m}_{1:B}=\{\boldsymbol{m}_1,\cdots,\boldsymbol{m}_B\}$, the pitch sequence is $p_{1:B}=\{p_1,\cdots,p_B\}$, and the loudness sequence is $l_{1:B}=\{l_1,\cdots,l_B\}$. 
Here $B$ denotes the number of frames. The inputs are transformed to sequences with the same dimension $\mathbb{R}^d$ by three MLP networks, respectively. These MLPs have the same structure, which contains three layers with a hidden 512 size. Then the outputs of MLPs are summed, which can be written as,
\begin{align}
    \boldsymbol{c}_{1:B} = f_1(p_{1:B}) + f_2(l_{1:B}) + f_3(\boldsymbol{m}_{1:B}),
\end{align}
where $f_1(\cdot)$, $f_2(\cdot)$, and $f_3(\cdot)$ denote functions of the three MLPs, respectively. The output $\boldsymbol{c}_{1:B}$ is transformed by a GRU \cite{gru} layer with 512 hidden size to $\boldsymbol{g}_{1:B}$. Subsequently, the combination of $\boldsymbol{g}_{1:B}$ and $f_1(p_{1:B})$ is mapped to a hidden sequence $\boldsymbol{\tilde{g}}_{1:B}$ by another MLP with three layers and $512$ hidden size. Finally, the hidden sequence $\boldsymbol{\tilde{g}}_{1:B}$ is projected to two sequences $\boldsymbol{\tilde{h}}_{1:B}$ and $\boldsymbol{\tilde{n}}_{1:B}$ by two distinct fully connected layers, respectively. These two sequences are upsampled with simple interpolation to harmonic content $\boldsymbol{h}_{1:T}$ and noise $\boldsymbol{n}_{1:T}$ whose length equals the length of the corresponding 8kHz audio. 

In order to evaluate the accuracy of the generated harmonic content and noise, we transform them to the 8kHz waveform by a reverb network, which is the same as the description in \cite{ddsp}. The spectral distance between the generated waveform and ground truth can be used to implement the backward propagation.

\subsubsection{BridgeNet}
\label{sec:bridgenet}

The generated harmonic content and noise by InstructNet are enough to synthesize the 8kHz audio. However, it cannot meet the demand of SVS tasks if we directly synthesize 48kHz audio from these features due to their rough periodic and aperiodic information. Therefore, we propose BridegNet to refine the information to be more sufficient and accurate. 
 
Specifically, the BridgeNet inherits the characteristics of UNet, further refining the harmonic content $\boldsymbol{h}_{1:T}$ and noise $\boldsymbol{n}_{1:T}$ generated by the InstructNet, as shown on the bottom middle of Figure \ref{fig:arch-instructsing}. The combination of $\boldsymbol{h}_{1:T}$ and $\boldsymbol{n}_{1:T}$ is first upsampled to a sequence $\boldsymbol{\tilde{u}}_{1:T'}$ whose length equals the length of the corresponding 48kHz audio by a transposed convlutional layer. Subsequently, $\boldsymbol{\tilde{u}}_{1:T'}$ is transformed to a latent variable sequence $\boldsymbol{u}_{1:T'}$ by the UNet \cite{ronneberger2015unet} module in BridgeNet, which can be formulated as
\begin{align}
    \boldsymbol{u}_{1:T'}=f_4(\boldsymbol{\tilde{u}}_{1:T'}),
\end{align}
where $f_4(\cdot)$ indicates the transformation function of the UNet module.

We argue that BridgeNet has the capability of improving the quality of periodic and aperiodic knowledge contained in the harmonic content and noise. The knowledge plays a key role in significantly accelerating the training speed of ExWaveNet, which will be illustrated in the next section.

\subsubsection{Extended WaveNet}
\label{sec:exwavenet}

The extended WaveNet \cite{hifiwavegan} is responsible for generating 48kHz high-fidelity singing voices according to the input mel-spectrogram and the refined latent variable sequence, as shown on the bottom right of Figure \ref{fig:arch-instructsing}. Concretely, the mel-spectrogram is upsampled to the length same with the latent sequence $\boldsymbol{u}_{1:T'}$. The combination of the upsampled output and the latent sequence is fed into the ExWaveNet module to generate audio. As for the detailed structure of ExWaveNet, we follow the design presented in \cite{hifiwavegan}. Compared with the original WaveNet \cite{oord2016wavenet}, ExWaveNet has a larger receptive field that can alleviate the glitches in the spectrogram of the generated 48kHz waveform.

\subsection{Discriminators}
\label{sec:discriminator}

For SVS tasks, periodic patterns and continuous long-term dependencies are crucial to distinguish the real and generated waveforms. To capture enriched features, we take advantage of two discriminators, as shown in the top right of Figure \ref{fig:arch-instructsing}. Firstly, we employed the multi-period discriminator (MPD) \cite{kong2020hifigan} to identify the periodic pattern from the time domain by reshaping the waveform according to the set of $M$ prime numbers in \cite{kong2020hifigan}. For each prime number, an individual sub-discriminator is used in the implementation.

Moreover, we enhanced the multi-resolution STFT discriminator (MRSD) \cite{univnet,safeear} by incorporating the multi-band analysis \cite{durian} to capture continuous long-term dependencies from the frequency domain \cite{normdetect}. The waveform is first converted to the frequency domain by applying STFT with various parameters, including (FFT size, frame shift, window length). In our implementation, we set these parameters to  (512, 128, 512), (1024, 256, 1024), (1024, 512, 1024), and (2048, 512, 2048). Subsequently, building upon the concept of multi-band analysis, these multi-resolution spectrograms are split equally into the low-frequency part, middle-frequency part, and high-frequency part. We employ an independent sub-discriminator for each sub-band of a specific STFT parameter, and there are $K$ sub-discriminators in total. In terms of the structure of the spectrogram discriminator for each sub-band, the Encodec architecture \cite{defossez2022highfi} is adopted in our implementation.

It is worth noting that the proposed MR-MBSD is distinct from the one in \cite{multibandmelgan}, which identifies the long-term dependencies from the time domain. We argue that it is easier to avoid the over-smoothing problem by utilizing the discriminator in the frequency domain, which is consistent with the conclusion in \cite{univnet}.

\subsection{Loss Function}
\label{sec:loss}

The final loss used in this paper to train InstructSing is a combination of multi-resolution spectrogram loss $\mathcal{L}_{sp}$ \cite{pwg} for generating realistic audio, mel-spectrogram loss $\mathcal{L}_{m}$ \cite{kong2020hifigan} for better auditory perception, which is applied to both 8kHz audio and 48kHz audio, feature match loss $\mathcal{L}_{fm}$ \cite{kumar2019melgan}, and adversarial loss $\mathcal{L}_{adv}$, as shown by the following formula.
\begin{align}
    \mathcal{L}_G  = \; & \lambda_1 * \mathcal{L}_{sp} + \lambda_2 * \mathcal{L}_{fm} + \\ \nonumber
    & \lambda_3 * (\mathcal{L}_{m}^{8k} + \mathcal{L}_{m}^{48k}) + \lambda_4 * \mathcal{L}_{adv}(G;D), \\
    \mathcal{L}_D  = \; & \mathcal{L}_{adv}(D;G), 
    \label{eq:final_gloss} 
\end{align}
where $\lambda_1$, $\lambda_2$, $\lambda_3$, and $\lambda_4$ are weights to balance the impact of these loss functions on the training process. They are set to 10, 1, 1, and 120, respectively.

We follow the implementation in \cite{pwg} to compute $\mathcal{L}_{sp}$, which encompasses spectral convergence and log STFT magnitude loss. For the feature match loss $\mathcal{L}_{fm}$, our approach adheres to the guidelines presented in \cite{kong2020hifigan, kumar2019melgan,li2022use}. This involves evaluating the L1 distance in feature maps of discriminators between real and generated audio. Furthermore, the quality of the generated mel-spectrogram is assessed by computing $\mathcal{L}_{m}$, which quantifies the L1 distance between the generated and real mel-spectrograms.

\begin{table*}[t]
\footnotesize
  \vspace{2mm}
  \setlength\tabcolsep{1.5pt}
  \centering
  \begin{tabular}{lccccccccc}
    \toprule
    \multirow{2}{*}{\textbf{Vocoder}} & \multirow{2}{*}{\textbf{RTF}($\downarrow$)} & \multicolumn{6}{c}{\textbf{MOS}($\uparrow$)} & \multirow{2}{*}{\textbf{STOI} ($400$k)($\uparrow$)} & \multirow{2}{*}{\textbf{PESQ} ($400$k)($\uparrow$)} \\
    \cmidrule(l{0em}r{0em}){3-8}
    & & $10$k & $20$k & $50$k & $100$k & $400$k & Opencpop($400$k) & & \\
    \midrule
    Ground truth (GT)  & -  & \multicolumn{5}{c}{$4.30\pm0.04$} & $4.33\pm0.04$ & -  \\
    HN \cite{ddsp} & $\boldsymbol{0.013}$ & $2.52\pm0.13$ & $2.77\pm0.12$ & $2.83\pm0.12$ & $3.02\pm0.10$ & $3.37\pm0.08$ & $2.98\pm0.10$ & $0.8306$ & $3.38$ \\
    RefineGAN \cite{refinegan} & $0.034$ & $3.11\pm0.09$ & $3.30\pm0.07$ & $3.55\pm0.07$ & $3.84\pm0.06$ & $4.11\pm0.06$ & $3.97\pm0.07$ & $0.9432$ & $4.01$ \\
    HN-uSFGAN \cite{hn-usfgan} & $0.070$ & $3.54\pm0.07$ & $3.77\pm0.06$ & $3.83\pm0.07$ & $4.05\pm0.07$ & $4.15\pm0.04$ & $3.99\pm0.06$ & $0.9489$ & $4.05$  \\
    InstructSing & $0.026$  & $\boldsymbol{4.05}\pm0.06$ & $\boldsymbol{4.15}\pm0.05$ & $\boldsymbol{4.17}\pm0.05$ & $\boldsymbol{4.20}\pm0.04$ & $\boldsymbol{4.25}\pm0.04$ & $\boldsymbol{4.17}\pm0.04$ & $\boldsymbol{0.9544}$ & $\boldsymbol{4.10}$ \\
    \bottomrule
  \end{tabular}
  \caption{Subjective and objective test result of the ground truth and generated audios of different vocoders for 48kHz singing voice synthesis. The MOS score was computed with a 95\% confidence interval.}
\label{tab:result_mos_svs}
\end{table*}

The loss function used in LS-GAN \cite{mao2017squares} is employed to alleviate the gradient vanishing in the training stage. It can be formulated as
\begin{align}
    \label{eq:adv_loss}
    \mathcal{L}_{adv}(G;D) = \; & \mathbb{E}_{\boldsymbol{z} \sim \mathcal{N}(0,1)}[(1 - D(G(\boldsymbol{z})))^2], \\
    \mathcal{L}_{adv}(D;G) = \; & \mathbb{E}_{\boldsymbol{y} \sim p_{data}}[(1 - D(\boldsymbol{y}))^2] + \nonumber \\ 
    & \mathbb{E}_{\boldsymbol{z} \sim \mathcal{N}(0,1)}[D(G(\boldsymbol{z}))^2],
\end{align}
where $G$ and $D$ represent the generator and discriminators in this paper, respectively, and $\boldsymbol{z}$ denotes the random noise input, while $\boldsymbol{y}$ is the singing voice from humans.

\section{Experiment}
\label{sec:exp}

\subsection{Dataset}
\label{sec:dataset}

All experiments were conducted using our internal 48kHz singing dataset, which comprises 16 male and 16 female singers. This dataset consists of 21,859 segments, with durations ranging from 4 seconds to 10 seconds. We randomly partitioned the dataset, reserving 400 segments for validation and 400 segments for testing. The remaining portion, totaling approximately 42 hours of audio data, served as the training dataset.

For the input feature, we applied a 1024-point STFT with a window length of 20ms and a frame shift of 5ms, resulting in a 120-dimensional mel-spectrogram by mel filterbank, which was subsequently normalized. In addition, pitch, loudness, and an 8kHz instructive signal were also extracted from the original data.

Furthermore, to assess the generalization capability of InstructSing on unseen singers, we conducted evaluations on the open-sourced Opencpop dataset \cite{opencpop} by randomly selecting 200 segments from this dataset as an additional test dataset.

\subsection{Experimental settings}
For the concrete structure of the model implemented in experiments, we configured the size of the hidden feature in InstructNet to 512. While for the BridgeNet, downsampling and upsampling rates were set to (8, 2, 2) and (2, 2, 8). Finally, the ExWaveNet was configured according to the description in \cite{hifiwavegan}, in which an 18-layer one-dimensional CNN with large kernel sizes was adopted to increase the size of the receptive field for generating waveform with better continuity.

InstructSing was trained using the AdamW optimizer \cite{adamw} with $\beta_1$ = 0.8, $\beta_2$ = 0.99, and weight decay $\lambda$ = 0.01. The learning rate was first increased by the warmup strategy for 5,000 steps from 0 to 0.0002 and subsequently decayed with a 0.999 factor in every iteration. To evaluate the inference speed of the model, we conducted the test on a single NVIDIA V100 GPU machine. Additionally, all of our testing work is based on 32-bit floating-point numbers and has not been quantified.

\begin{figure}[t]
  \centering
  \includegraphics[width=1\linewidth]{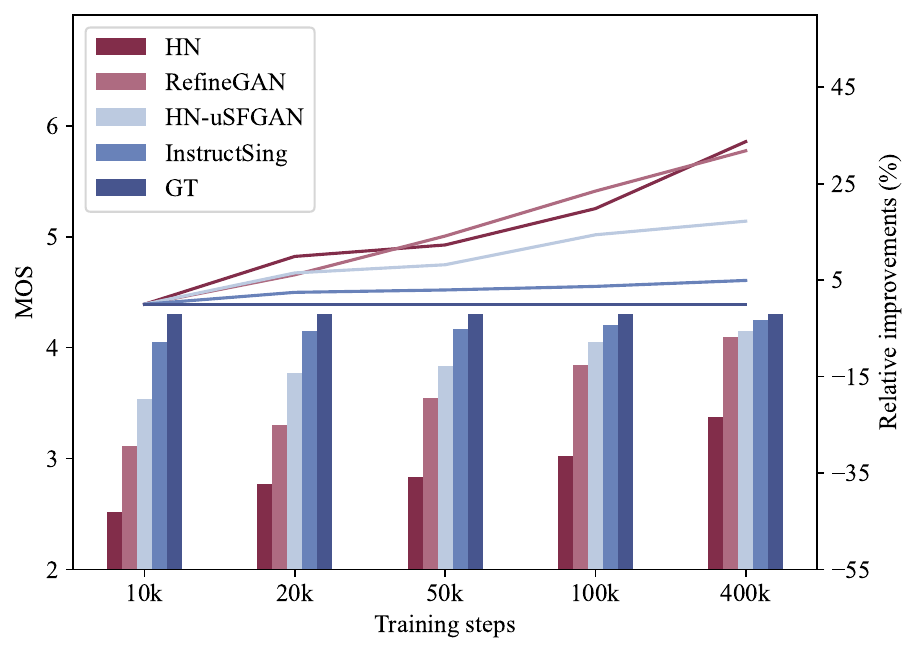}
  \caption{The variation of MOS at different training steps. The left vertical axis and bar graphs indicate the absolute value of the MOS score for each vocoder. The right vertical axis and line graphs represent relative improvement based on each vocoder's MOS score at the $10$k step.}
  \label{fig:training_mos}
\end{figure}

Given InstructSing's fusion of DDSP with adversarial training, our comparative analysis involves benchmarking against two key models: the DDSP-based harmonic-plus-noise (HN) model \cite{ddsp} and RefineGAN, a GAN-based SVS vocoder \cite{refinegan}. 
Moreover, acknowledging the sine excitation-like role of the latent variable sequence in our proposed model, it becomes essential to contrast InstructSing with a model utilizing a similar sine excitation signal to expedite training. To fulfill this requirement, we've opted for HN-uSFGAN \cite{hn-usfgan}, recognized for its strong performance, as documented in \cite{hn-usfgan}. This comparative study allows us to gauge InstructSing's efficacy within a context where models leverage akin structural components to drive their training processes.

All vocoders used in experiments were combined with the acoustic model XiaoiceSing2 \cite{xiaoice2}, which is a high-fidelity singing voice synthesizer responsible for predicting 120-dimensional mel-spectrogram, pitch sequence, and loudness sequence.

\subsection{Experimental result}
\label{sec:analysis}

In terms of the experimental result, we pay attention to the inference speed and generated audio quality due to their importance for a vocoder. As shown in Table \ref{tab:result_mos_svs}, the real-time factor (RTF) was computed for each vocoder. DDSP-based HN model achieved the fastest inference speed since it is a lightweight vocoder compared with others \cite{ddsp}. Although the inference speed of InstructSing is slower than the HN model, it is acceptable for a high-quality neural vocoder in practice. Moreover, InstructSing was faster than other GAN-based neural vocoders, such as RefineGAN and HN-uSFGAN shown in Table \ref{tab:result_mos_svs}.

As for the audio quality, we conducted subjective and objective evaluations on ground truth and synthesized singing voices. Mean Opinion Score (MOS) was adopted as the metric in the subjective evaluation, which involved preparing 50 segments of singing voices for each vocoder and ground truth. Thirty listeners were employed to participate in the test. In Table \ref{tab:result_mos_svs}, we calculate the MOS for all vocoders at different training steps, including 10k, 20k, 50k, 100k, and 400k. Regardless of the training step, InstructSing can achieve the highest MOS score in all neural vocoders. This can also be reflected in the bar graphs in Figure \ref{fig:training_mos}. As the number of training steps increases, the MOS is improved for all vocoders, which meets our expectations. However, suppose we calculate the relative improvements for each vocoder at other training steps based on its MOS score at the 10k step. In that case, we can find that InstructSing achieves the smallest relative enhancements as line graphs show in Figure \ref{fig:training_mos}, which means that InstructSing can converge well within much fewer training steps. For example, the difference between MOS at 20k and 400k training step is 0.1 for InstructSing, which is much smaller than the difference of HN-uSFGAN.

In addition to the evaluation of in-domain data, we also evaluate our model on the test data from an unseen singer of the Opencpop dataset \cite{opencpop}. All vocoders trained for 400k steps are evaluated on this testing dataset. From Table \ref{tab:result_mos_svs}, it is obvious that InstructSing has the smallest MOS difference between the unseen singers and seen singers. It indicates that when synthesizing voices for unseen singers,  our proposed InstructSing has better generalization ability than other vocoders.

For objective evaluation, we calculated the STOI and PESQ scores for all vocoders, and the results are shown in Table \ref{tab:result_mos_svs}. From these numbers, the performance of InstructSing is clearly the best among all vocoders. This further proves that in the case of 48kHz SVS, InstructSing is more trustworthy compared to other vocoders.

\begin{table}[t]
  \vspace{4mm}
  \setlength\tabcolsep{10pt}
  \centering
  \begin{tabular}{lc}
    \toprule
    \textbf{Vocoder}   & \textbf{MOS}($\uparrow$)  \\
    \midrule
    Ground truth                                & $4.30\pm0.04$     \\
    InstructSing-8K                             & $4.25\pm0.04$  \\
    \midrule
    InstructSing-4K                             & $3.97\pm0.07$  \\
    InstructSing-16K                            & $4.10\pm0.05$    \\
    InstructSing-24K                            & $4.08\pm0.06$     \\
    \midrule
    w/o Multi-Band                              & $4.05\pm0.05$ \\
    Pulse Extractor \cite{hifiwavegan}          & $3.85\pm0.08$ \\
    \bottomrule
  \end{tabular}
  \caption{Ablation study to show MOS score at $400$k step of different configurations, including sampling rate of instructive signal, w/o multi-band, and Pulse Extractor.}
\label{tab:result_mos_svs_ab}
\end{table}

\subsection{Ablation study}
\label{sec:ablation}

In this paper, we utilized DDSP to generate an instructive signal to guide the following adversarial training. It is necessary to test which sampling rate is suitable for the instructive signal. In addition to the 8kHz sampling rate used in the primary experiment, we conducted an ablation experiment for other sampling rates, including 4kHz, 16kHz, and 24kHz, as shown in Table \ref{tab:result_mos_svs_ab}. 

It can be seen that there is a significant decrease in the MOS score of InstructSing-4k compared to InstructSing-8k since it has obvious electrical sound according to the feedback from listeners. This may be due to the lack of the necessary details in the instructive signal with the lower sampling rate.

For InstructSing-16k and InstructSing-24k, although the MOS scores do not show as significant a decrease as InstructSing-4k, they still decrease by 0.15 and 0.17, respectively. This may be attributed to the disability of DDSP in synthesizing audio with a high sampling rate because it will introduce abnormal harmonic information into the signal. Overall, the results indicate that selecting an appropriate sampling rate is crucial for achieving high-quality singing voice synthesis.

Additionally, we also conducted an ablation study on the discriminator. We used the MRSD proposed in \cite{univnet} to replace our proposed MR-MBSD, and the results showed that the MRSD was slightly worse (-0.2) than that of MR-MBSD in terms of the MOS score. It indicates that multi-band analysis is beneficial to improving the performance of the generator.

Finally, to determine the role of InstructNet and BridgeNet, we also conducted an ablation study by replacing them with the Pulse Extractor presented in \cite{hifiwavegan} to generate the pulse sequence for ExWaveNet. It is not difficult to see from Table \ref{tab:result_mos_svs_ab} that the MOS score differs by 0.4 compared to the result of InstructSing-8k, which fully proves the role of InstrcutNet and BridgeNet in InstructSing.

\section{Conclusion}
\label{sec:conclusion}

In this paper, we introduce a new high-fidelity vocoder for the task of singing voice synthesis, namely InstructSing. It combines DDSP with adversarial training by connecting InstructNet and ExWaveNet via BridgeNet. The InstructNet first generates the harmonic and noise, which are enough to synthesize the audio with a low sampling rate. Subsequently, harmonic and noise are refined by the BridgeNet to generate a latent variable sequence containing sufficient periodic and aperiodic information as an instructive signal. Finally, the ExWaveNet can synthesize 48kHz audio by utilizing the latent variable sequence and mel-spectrogram. Additionally, we also improve the multi-resolution STFT discriminator by incorporating multi-band analysis into it. Through our experiments, it can be seen that the InstructSing model can converge much faster compared with other SOTA neural vocoders and the singing voice generated by it achieves human-level quality after sufficient training iterations in terms of the MOS metric. In the future, we aim to further optimize the inference speed of InstructSing on CPU-only machines.

\section{Ethics Statement}
This research on InstructSing, a novel high-fidelity singing voice generation model, is firmly rooted in ethical principles pertinent to AI and voice synthesis technology. Acknowledging the critical importance of respecting individual privacy, we ensure strict adherence to data protection norms and uphold the necessity of informed consent. In constructing and utilizing our datasets, we pay careful attention to preventing bias, thereby fostering fairness and inclusivity in representing diverse vocal attributes. We recognize and address the inherent risks of misuse associated with voice synthesis technology, advocating for responsible application and maintaining transparency in our methods and processes. Our commitment lies in advancing technology ethically, ensuring that our contributions to the field not only push the boundaries of innovation but also consider the broader societal and individual implications.

\bibliographystyle{IEEEbib}
\bibliography{strings,refs}

\end{document}